\documentclass[journal, draftclsnofoot, onecolumn]{IEEEtran}
\usepackage{cite}

\usepackage{algorithm}
\usepackage{algpseudocode}
\usepackage{graphicx}
\usepackage{amsmath}
\usepackage{amssymb}
\usepackage{relsize}
\usepackage{multirow}
\usepackage{xcolor}
\usepackage{bbm}

\begin{document}
\bstctlcite{IEEEexample:BSTcontrol}
\title{Resource Provisioning in Edge Computing for Latency Sensitive Applications}

\author{Amine~Abouaomar,~\IEEEmembership{Student~Member,~IEEE,}
       Soumaya~Cherkaoui,~\IEEEmembership{Senior~Memeber,~IEEE,}
       Zoubeir~Mlika,~\IEEEmembership{Memeber,~IEEE,}
        and~Abdellatif~Kobbane,~\IEEEmembership{Senior~Memeber,~IEEE}
\thanks{A. Abouaomar is with INTERLAB, Engineering Faculty, Universit\'e de Sherbrooke, Canada and also with ENSIAS, Mohammed V University in Rabat, Morocco (e-mail: amine.abouaomar@usherbrooke.ca)}
\thanks{S. Cherkaoui is with INTERLAB, Engineering Faculty, Universit\'e de Sherbrooke, Canada.}
\thanks{Z. Mlika is with INTERLAB, Engineering Faculty, Universit\'e de Sherbrooke, Canada.}
\thanks{A. Kobbane is with the ENSIAS, Mohammed V University in Rabat, Morocco.}
}

\markboth{}%
{Abouaomar \MakeLowercase{\textit{et al.}}: Resource Provisioning in Edge Computing for Latency Sensitive Applications}

\maketitle

\begin{abstract}
Low-Latency IoT applications such as autonomous vehicles, augmented/virtual reality devices and security applications require high computation resources to make decisions on the fly. However, these kinds of applications cannot tolerate offloading their tasks to be processed on a cloud infrastructure due to the experienced latency. Therefore, edge computing is introduced to enable low latency by moving the tasks processing closer to the users at the edge of the network. The edge of the network is characterized by the heterogeneity of edge devices forming it; thus, it is crucial to devise novel solutions that take into account the different physical resources of each edge device. In this paper, we propose a resource representation scheme, allowing each edge device to expose its resource information to the supervisor of the edge node through the mobile edge computing application programming interfaces proposed by European Telecommunications Standards Institute. The information about the edge device resource is exposed to the supervisor of the EN each time a resource allocation is required. To this end, we leverage a Lyapunov optimization framework to dynamically allocate resources at the edge devices. To test our proposed model, we performed intensive theoretical and experimental simulations on a testbed to validate the proposed scheme and its impact on different system’s parameters. The simulations have shown that our proposed approach outperforms other benchmark approaches and provides low latency and optimal resource consumption. 
\end{abstract}


\IEEEpeerreviewmaketitle

\section{Introduction}

\begin{table}[h]
	\centering
	\caption{Summary of System Variables}
	\begin{tabular}{l l}
		\hline \hline
		Symbol & Description \\ \hline
		$\mathcal{U}$ & Set of users \\ \hline
		$\mathcal{E}$ & Set of EN \\ \hline
		$x^{u}_{i}$ & Association index \\ \hline
		$D_i$ & Devices of the EN i \\ \hline
		$R_{(i,j)}$ & Set of resources of device j \\ \hline
		$\mathcal{C}_{(i,j)}$ & Set of containers of device j \\ \hline
		$S_{(i,j)}$ & Set of services \\ \hline
		$V^{s_{(i,j)}}_{r}$ & Requirement resource vector a service \\ \hline
		$\mathbbm{1}_r$ & Identification function \\ \hline
		$A_{s_{(i,j)}}$ & Matrix of required resources per service \\ \hline
		$T_{out}$ & Timeout of a request \\ \hline
		$\varrho_{(i,j)}$ & Request \\ \hline
		$\Xi^{\sim,p}_{(i,j)}$ & Amount of resources consumed by a container \\ \hline
		$\lambda_i$ & Arrival rate of requests \\ \hline
		$Rq_i$ & Number of request at a time slot \\ \hline
		$\delta^{\sim}_{L}$ & Delay for a given resource \\ \hline
		$D_{(i,j,l)}$ & Data size for local processing \\ \hline
		$f_\mathcal{P}$ & Processing capacity \\ \hline
		$idx_{r/w}$ & Read/Write index \\ \hline
		$\xi_{i}^{u}$ & Link data rate \\ \hline
		$h_.$ & Channel gain \\ \hline
		$P_.$ & Power of communication \\ \hline
		$\rho_{\sim}^{i}$ & Data size for edge processing \\ \hline
		$Q_i$ & Queue vector of an EN i \\ \hline
		$\mathcal{H}$ & History vector \\ \hline
		$b_i$ & Queue's request arrival process \\ \hline
		$a_i$ & Queue's service arrival process \\ \hline
		$Z_i$ & Queue dynamic of EN i \\ \hline
		$y_k$ & Penalty process variable \\ \hline
		$L^{r^{\sim}_{(i,j)}}_{(i,j,k)}$ & Load for a given resource \\ \hline
		$V$ & Trade off parameter \\ \hline
		$p^{avg}_{m}$ & average resource loss \\ \hline
		$\alpha_i(t)$ & Resource allocation scheme \\ \hline
		$\beta_i(t)$ & Resource state vector \\ \hline
		$y_i(t)$ & Vector of penalties \\ \hline
		$L^{norm}_{(i,j,k)}$ & Normal load \\ \hline
		$rate_{norm}$ & Rate of request arrival \\ \hline
		\hline
	\end{tabular}
\end{table}
The evolution of technologies and communication systems in the last decade gave birth to new Internet-based applications and services that require low latency. Cloud computing was proposed as a powerful technology to enable low latency requirements and optimized resource consumption by offering many advantages such as high availability, scalability and reduced costs \cite{Era_Perso_Cloud, 7835337}. However, to meet the latency requirement in tasks processing, remote cloud computing infrastructure may not be suitable for latency-sensitive applications such as industrial process monitoring, automated vehicles \cite{7585028, 9497103}, virtual/augmented reality, surveillance and security and human emotions detection. Therefore, the proximity to the processing infrastructure is the key when it comes to reducing the experienced latency \cite{7807196, 9318243}. Such infrastructure is located at the edge of the network, enabling to address the limitations of cloud computing infrastructure through distributed and low latency computation.

Edge computing (EC) has emerged as a promising solution to lower the experienced latency by distributing the processing, communication and control closer to where the data is generated \cite{9240934, 7807196, tocze2018taxonomy}. EC therefore extends cloud computing by providing applications with the computational, storage and communication resources at the edge of the network. The main standardization efforts in EC were initially proposed by the Industry Specification Group (ISG) within the European Telecommunications Standards Institute (ETSI) \cite{ETSIGRMEC027,GSMEC010,GSMEC011,GSMEC021,ETSI5G}. Unlike cloud computing, which is scalable and highly maintained, EC is more heterogeneous and resource constrained. The edge of the network is built-up on edge nodes (ENs). ENs are the aggregation of heterogeneous edge devices (EDs) such as edge routers and switches, edge servers, sensors and even some end users (EUs) equipment (e.g. laptops and smartphones). From the location point of view, the EDs are located outside the premises of the operator which makes the maintenance and management difficult. From the architectural point of view, EDs could have different hardware architectures and have different capabilities. For instance, an EN could have several edge routers and edge servers. Some edge servers may have high processing resources or specialized ones such as graphical processing units (GPUs), that make them suitable to perform intense computation tasks. Some other edge servers could have big storage space to serve as caching and storing infrastructure at the edge of the network \cite{8238174, 7451194}. Finally, edge routers may have moderate processing resources but powerful networking resources, allowing them to handle more traffic than others.

The disparity of available resources at the edge of the network requires an efficient and optimal resource provisioning, especially because of the limited available resources compared to the cloud. However, an adequate and optimal resource provisioning requires a good knowledge of available resources. The unavailability of information about EDs capabilities at the ENs can increase the latency and cause additional delay because some of EDs could be asked to perform tasks they are noted best suited for. Therefore, providing the information about ED's resources at an EN will make the process of tasks distribution more optimal. Thus, any supervising entity that oversees tasks distribution can build suitable resource provisioning schemes based on this information.

In this paper, we propose a resource representation model to characterize the different physical resources of the EDs. There are four resources of interest in this paper: processing, storage, memory and networking. Since each ED is aware of its different available resource, the ENs will be capable of discovering the resource information of the EDs independently of the architecture or manufacturer. In the literature, most of the works investigate the problem of edge resource allocation considering only communication and computation resources \cite{8962353,liu2019dynamic,8683499,guo2019mobile,8538500,8815852,liao2019learning,zhou2019computation,gu2019task}. A few works also consider other types of resources but without specifically targeting the delay minimization problem \cite{arkian2017mist,skarlat2017towards,yousaf2016fine}. In this work, we formulate the resource allocation optimization problem as a Lyapunov optimization with the aim of minimizing the overall applications experienced latency by jointly considering all the above-mentioned resources. The main contributions of this paper re summarized as follows:

\begin{itemize}
	\item To characterize the physical resource in an EC environment, we propose a resource representation model which adheres to the ETSI standard \cite{ETSIGRMEC027,GSMEC010,GSMEC011,GSMEC021,ETSI5G}. The model represents different kinds of resources (processing, storage, memory and networking) of EDs and their capabilities.
	\item We propose a resource allocation scheme based on Lyapunov optimization to minimize the experienced delay through the study of the service queues dynamics at the ENs.
	\item We propose an algorithm to optimize the frequency of resource allocation and information exchange. The proposed algorithm is based on the workload at each EN.
	\item We simulate the proposed scheme under different configurations of parameters. We also use a testbed inspired by the work in \cite{rimal2018experimental} to experiment the proposed scheme. We evaluate the latency, the consumption of different resources at the EDs and the queue evolution over time. The simulations show that our proposed approach outperforms other benchmark approaches and provides low latency and optimal resource consumption.
\end{itemize}

The reminder of the paper is organized as follows. The system model is first detailed in Section II. Subsequently, we formulate the problem of latency minimization, the dynamic of the queues and the different requirements in Section III. In Section IV, we detail the proposed resource representation scheme, the resource allocation scheme and the task distribution model. The performance of the system is simulated in Section V. We discuss the related works in Section VI. And we conclude the paper in Section VII.

\section{System Model}
\begin{figure*}
	\centering
	\includegraphics[width=0.7\linewidth]{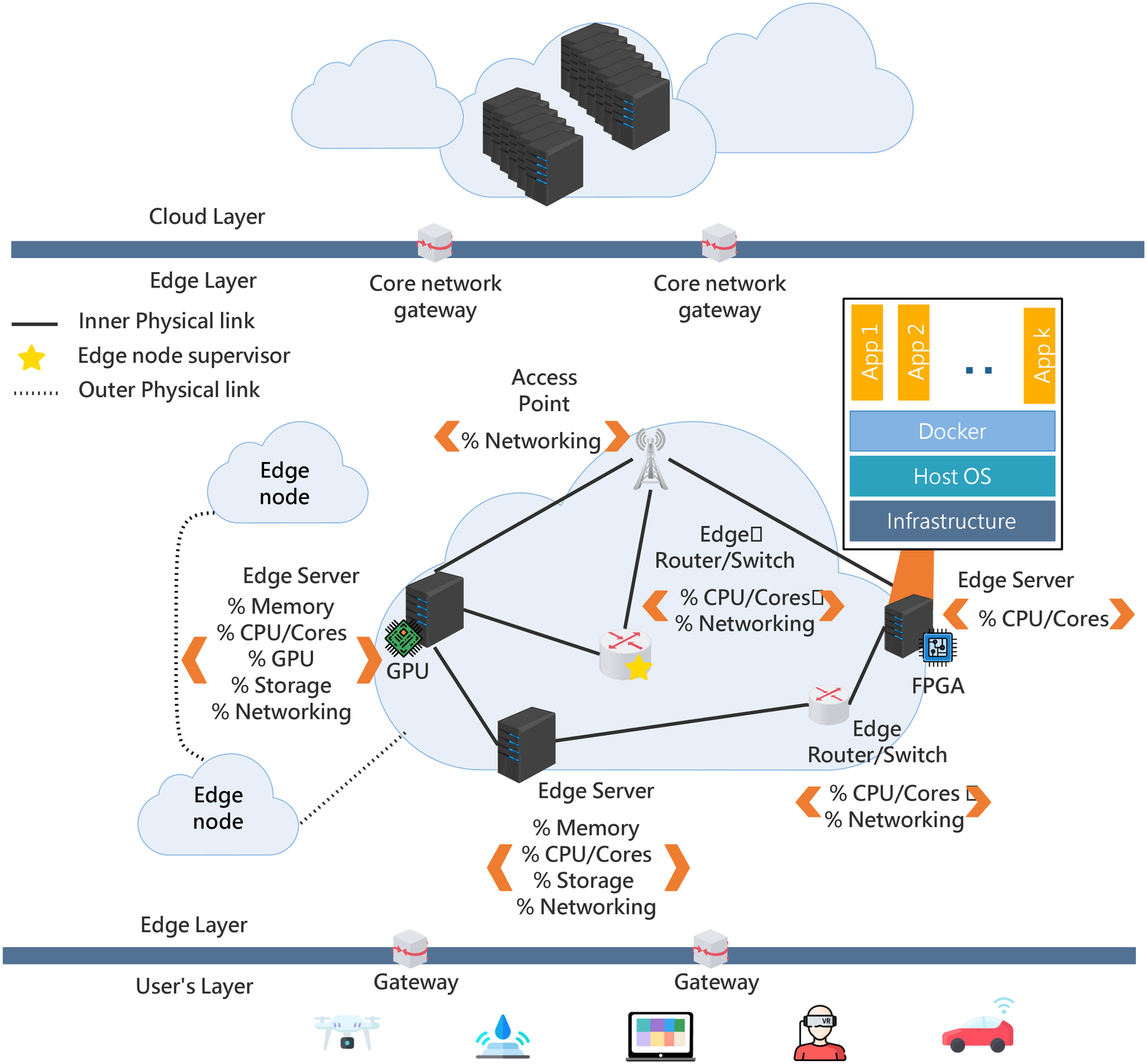}
	\caption{Overview of the different entities of the edge computing architecture.}
	\label{fig:edge-arch}
\end{figure*}
\subsection{Network architecture}
Let us consider an edge computing network architecture as the one depicted in Fig. \ref{fig:edge-arch} that adheres to the requirement of ETSI standard based architecture \cite{ETSIGRMEC027, GSMEC010, GSMEC011, GSMEC021, ETSI5G}. In this architecture, the end users are depicted in the lower layer which is called EUs' layer. EUs can be drones, surveillance cameras or virtual reality equipment. For simplicity, EU are to be referred to as users.
Let $\mathcal{U}$ be the set of $N$ users $\mathcal{U} = \{u_1, ..., u_N\}$. Users communicate with the edge layer through gateways that connect to one or several edge nodes (ENs). The gateways are used as intermediary devices to connect the user equipment to the edge nodes. For example, these intermediary devices can be routers or switches within the same level of the hosting organization. ENs belong to the set $\mathcal{E} = \{e_1, e_2, . . . , e_M\}$ of $M$ EN. We assume that users are already associated to ENs following the strategy proposed in \cite{8647545}. Let $x_i^u(t)$ be a binary variable that ensures the user-EN association, described as follows:
\begin{equation}
	\label{eq:assoc}
	\begin{cases}
		x_{i}^{u}(t) \in \{0,1\}, \forall i\in \mathcal{E}, \forall u \in \mathcal{U} \\
		\mathlarger{\sum_{i\in\mathcal{E}} x_{i}}^{u}(t) = 1, \forall u \in \mathcal{E}
	\end{cases}
\end{equation}
Each EN is considered to be an aggregation of heterogeneous EDs denoted by $D_i = \{d_{(i,1)}, d_{(i,2)}, ... , d_{(i,K)}\}$. In addition, users-EN matching is distributed which makes the system standalone, but in case we have a huge number of EDs, the discovery phase could take much more time. Therefore, each EN $e_i$ is supervised by an edge node supervisor (ENS), denoted $e_i^*$ . Each ED has resources in terms of processing ($\mathcal{P}$), storage ($\mathcal{S}$), memory ($\mathcal{M}$) and networking ($\mathcal{N}$), denoted by the set $R_{(i,j)} = \{r_{(i,j)}^{\sim} , \sim \in \mathcal{P}, \mathcal{S}, \mathcal{N}, \mathcal{M}\}$, with $i$ representing the EN and $j$ the ED. Each ED $j$ hosts a set of containers \cite{merkel2014docker} $C_{(i,j)} = \{c_{(i,j)}^{1}, ..., c_{(i,j)}^P\}$ representing the different services deployed on the EN $i$. We denote by $\Xi_{(i,j)}^{\sim,k}$ the amount of resource $\sim$ consumed by the container $k$ executed on the ED $j$ at the EN $i$. Let $S_{(i,j)} = \{s_{(i,j)}~|~i~\in~\mathcal{E}~and~j~\in~D_i\}$ be the set of services hosted on the $j^{th}$ ED at the $i^{th}$ EN. Each service requires resources in terms of processing, storage, memory and networking, which are represented by a vector $V_{r}^{s_{(i,j)}} = \langle \mathbbm{1}_r, A_{s_{(i,j)}}\rangle$ where $\mathbbm{1}_{r}(\sim) = 1$ if the resource $\sim$ is required by $s_{(i,j)}$ and $\mathbbm{1}_{r}(\sim) = 0$ otherwise. $A_{s_{(i,j)}}$ represents the resource amounts required by service $s_{(i,j)}$. The resource capacity at each device is a crucial parameter that should be considered for a good resource allocation. The following equation describes the constraint on the resource capacities and is given as:
\begin{equation}
	\label{eq:resconc}
	\sum_{k=1}^{|C_{(i,j)}|} \Xi_{(i,j)}^{\sim,k} \leq r_{(i,j)}^\sim ~|~ \sim~\in \{\mathcal{P}, \mathcal{M}, \mathcal{N}, \mathcal{S}\}
\end{equation}
$C_{(i,j)}$ represents the set of containers, $\Xi_{(i,j)}^{(\sim,k)}$ represent the amount of resources required by the container $k$, $r_{(i,j)}^{~}$ represents the maximum capacity of the resource $\sim$.
Users perform service requests on the EN. We denote a single request by $\rho_{(i,j)} = \langle D_{(i,j,l)}, s_{(i,j)}, T_{out}\rangle$, where $D_{(i,j,l)}$ is the data size to be processed by service $s_{(i,j)}$ and $T_{out}$ is the time duration upon which the user gives up service $s_{(i,j)}$. Without loss of generality, the time is considered to be discreet and indexed by slots $t~\in~\mathbb{N}$ and the requests are identically independent distributed (i.i.d.). The requests arrival rate is then proportional to the number of request, and for a given EN i we have the arrival rate \cite{liu2019dynamic}:
\begin{equation}
	\lambda_{i} = \frac{\mathbb{E} \left[ V_{r}^{s_{(i,j)}} \right]}{t}
\end{equation}
where $V_{r}^{s_{(i,j)}}$ represents the vector of the required resources by the service $s_{(i,j)}$ at the time slot $t$.
In this paper, we consider that requests are either processed locally, using the residual resources of the user, processed at the edge, or both in a proportional manner; locally and at the EN.

\subsection{Local processing}
Local processing delay consist in both computational and storage delay. The computational delay is given as follows:
\begin{equation}
	\delta_L^\mathcal{P} = \frac{D_{u_{k}}}{f_\mathcal{P}}
\end{equation}
where, $f_\mathcal{P}$ represents the computing capacity of the ED which is the amount of data that the processing unit can process per time unit. $D_{uk}$ represents the data size. At this stage there is no communication. Therefore, no queuing delay which represents the time that the request spends at the queue of a shared resources (the EN in this case). For the local processing case, the user equipment uses its own capabilities to process the task. Thus, the communication delay is, $\delta_L^C = 0$.
We considered the storage delay which is the delay from storing/reading to/from the storage support of ED. Since the ED are heterogeneous, some of the EDs can be equipped with HDD storage supports while others might have SSD, and it is well known that writing/reading to/from SSD is faster than HDD \cite{rydning2018digitization, 8360858}. In addition the caching at the edge represents a use case in which the storage performance have an important impact on the latency \cite{8491367}. For these considerations we considered the storage delay, and we put:
\begin{equation}
		\delta_L^\mathcal{S} = D_{u_k}\left(\frac{idx_w}{s_w} + \frac{idx_r}{s_r}\right)
\end{equation}
where, $s_w$ and $s_r$ are respectively the storage support writing and reading speed of the ED, $idx_w$ and $idx_r$ are binary variables denoting either writing/reading is active. Reading index is defined as follows:
\begin{equation}
	idx_r = \begin{cases}
		0 \text{ if request requires no reading} \\
		1 \text{ otherwise}
	\end{cases}
\end{equation}
and writing index as:
\begin{equation}
	idx_r = \begin{cases}
		0 \text{ if request requires no writing} \\
		1 \text{ otherwise}
	\end{cases}
\end{equation}

In this case of local processing, the total delay experienced by a given request is given as:
\begin{equation}
\delta_{L}^{total} = \delta_{L}^\mathcal{P}+\delta_{L}^\mathcal{S}
\end{equation}
In this paper, the queuing delay represents the time that the request spends at the queue of a shared resources in the edge network. For this reason, we considered the queuing delay in the edge processing scenario and we adopted the Lyapunov framework to study the queues dynamics. For the local processing case, the queuing delay is not considered, because the user equipment uses its own capabilities to process the task and thus it has no queue of a shared resources.
\subsection{Edge processing}

We consider that the links between the users and EN are reliable, also, the communication setup is already done. Processing at the EN entails a queuing delay and a propagation delay. The propagation delay is the sum of, (i) the delay that a service request experiences to get to the EN and (ii) the delay $t$ that the corresponding response experiences to get back to the user from the EN. This propagation delay depends on the communication link’s data rate, which is defined as follows \cite{liu2019dynamic}:

\begin{equation}
\xi_i^u = B_w.\log_2\left(1 + \frac{P_u\left|h_{(u,i)}\right|^2}{\sigma^2 + \mathlarger{\sum_{\substack{v\in\mathcal{U}\\v\neq u}}} {P_v\left|h_{(v,i)}\right|^2}}\right),
\end{equation}
where $B_w$ represents the bandwidth of the communication link, $P_u$ is the transmission power or user $u$, $h_{(u,i)}$ is the channel gain between user $u$ and EN $i$ and $\sigma^2$ represents the noise variance. The transmission is the ratio of the packet size to be transmitted to the communication data rate and is given as follows \cite{9383093}:
\begin{equation}
\delta_{(i,u)}^\mathcal{C} = \frac{\rho_i^\sim}{\xi_i^u}
\end{equation}
The total delay for task processing is given as:
\begin{equation}
\delta_{(i,u)}^{total} = \delta_{(i,u)}^{\mathcal{C}} +  \delta_{(i,u)}^{\mathcal{S}} +  \delta_{(i,u)}^{\mathcal{P}} +  \delta_{(i,u)}^{wait}
\end{equation}
where $\delta_{(i,u)}^{\mathcal{\mathcal{C}}}$, $\delta_{(i,u)}^{\mathcal{S}}$, $\delta_{(i,u)}^{\mathcal{P}}$ and $\delta_{(i,u)}^{wait}$ represent respectively, the transmission delay, storage delay, processing delay and the waiting time at the queue of the EN.

\subsection{Problem Formulation}

In order to optimize the total delay, we start by analyzing the queuing delay which is an essential component of the overall delay. To achieve that, we could use tools such as Little's Law that affirms that the queuing time is proportional to queue length. However, this would allow to consider the queue length and not the queue tail, which would not help guaranteeing the constraints regarding the low latency. We leverage the Lyapunov optimization framework due to its capabilities to provide optimized and stable queuing dynamic \cite{neely2012stability}. 

We assume that all the requests are stored in the same queue and i.i.d. Let $Q_i(t)$ be the queue vector of EN $i$ that evolves in time. The evolution of the queue $Q_i(t)$ is based on the event of requests arrival. Let us assume that the value of $Q_i(t)$ have the structure of $\{\mu(-1), \mu(0), \mu(1), ... \}$ where $Qi(0) = \mu(-1)$ is the initial state of the vector $Q_i$, where $\mu(t)$ represents the requests arrived in time $t$. The values of $Q_i(t)$ are based on the values of $\{Q(0), \mu(0), ... , \mu(t - 1)\}$ and let $\mathcal{H}(t)$ be the history vector up to the time $t - 1$. The queue $Q_i(t)$ evolves in time as follows:
\begin{equation}
	\label{eq:Q}
	Q_{i}(t+1) = \max\left[\left(Q_i(t)-b_i(t)+a_i(t)\right), 0\right]
\end{equation}

Where $a_i(t)$ and $b_i(t)$ represents respectively the arrival process and the service process \cite{neely2012stability}. Values of $a_i(t)$ and $b_i(t)$ are defined by general functions $\hat{a_i}(t)$ and $\hat{b_i}(t)$ respectively ads defined in Eq. (\ref{eq:genfunc}). In addition, we define the queue dynamic at the EN $i$ as follows:

\begin{equation}
	\label{eq:Z}
	\begin{split}
		Z_{i}(t+1) = \lim\limits_{t\to\infty}\frac{1}{t}\mathlarger{\sum_{\tau=1}^{t}}P(\max\left[\left(Q_i(t)+b_i(t)\right)-a_i(t), 0\right] \\
	 	> \Delta_{T_{out}}(t))
	\end{split}
\end{equation}

Where, $Q_i(t)$ is the queuing dynamic at the EN $i$, and $\Delta_{Tout}(t)$ is the queue length bounds when the tolerable bound relative to the timeout of the tasks. For a given resource allocation scheme $\alpha_i(t)$ and a resource state $\beta_i(t)$ of the EDs in the EN $i$. $\beta_i(t)$ is used as intermediary variable to find the optimal value of $\alpha_i(t)$ that will be used as input to our resource allocation algorithm. For each $\beta_i(t)$, which is the resource state represented as a vector with values of different values of available resources there is an associated $\alpha_i(t)$, which is the resource allocation scheme.

The objective is to find an optimal value of $\alpha_i(t)$. Each allocation scheme $\alpha_i(t)$ incurs a vector of penalties $y_i(t) = \{y_0(t), y_1(t), ... , y_K(t)\}$.
The arrival process $a_i(t)$, the service process $b_i(t)$, and the penalty vector $y_i(t)$ can be expressed as a function of $\alpha_i(t)$ and $\beta_i(t)$ as follows \cite{neely2012stability}:

\begin{equation}
	\label{eq:genfunc}
	\begin{cases}
		a_i(t) = \hat{a_i}(\alpha_i(t), \beta_i(t))\\
		b_i(t) = \hat{b_i}(\alpha_i(t), \beta_i(t))\\
		y_k(t) = \hat{y_k}(\alpha_i(t), \beta_i(t))
	\end{cases}
\end{equation}

For $t > 0$, we define $\bar{a}_i(t)$, $\bar{b}_i(t)$, $\bar{y}_k(t)$ and $\bar{Q}_i(t)$ the average sizes of $a_i(t)$, $b_i(t)$, $y_k(t)$ and $Q_i(t)$ defined as :

\begin{equation}
	\label{eq:avgsize}
	\begin{cases}
		\bar{a}_i(t) = \frac{1}{t} \mathlarger{\sum_{\tau=0}^{t-1}}a_i(\tau) \\
		\bar{b}_i(t) = \frac{1}{t} \mathlarger{\sum_{\tau=0}^{t-1}}b_i(\tau)\\
		\bar{y}_k(t) = \frac{1}{t} \mathlarger{\sum_{\tau=0}^{t-1}}y_k(\tau)\\
		\bar{Q}_i(t) = \frac{1}{t} \mathlarger{\sum_{\tau=0}^{t-1}}Q_i(\tau)
	\end{cases}
\end{equation}
The ENS compute the optimal value of $\alpha_i(t)$ through solving the following optimization problem:
\begin{equation}
	\label{eq:limy}
	\lim\sup\limits_{t\to\infty}\bar{y}_0(t)
\end{equation}

Subject to 
\begin{itemize}
	\item (C1') : $\lim\sup\limits_{t\to\infty} \bar{y}_k(t)\leq0$
	\item (C2') : Stability of $Q_i(t)~\forall~t~\in\{0, 1, ...\}$
\end{itemize}

The delay optimization problem is formulated as follows:

\begin{equation}
	\label{eq:mindelta}
	minimize~\{\delta_{(i,u)}^{total}\}
\end{equation}
subject to:

\begin{itemize}
	\item (C1): Eq. (\ref{eq:resconc}) to ensure resource consumption and maximum capacities.
	\item (C2): Eq. (\ref{eq:assoc}) for user-EN association.
	\item (C3): Eq. (\ref{eq:Q}), (\ref{eq:Z}) to ensure the queue dynamic.
	\item (C4): Eq. (\ref{eq:genfunc}), (\ref{eq:avgsize}) for resource allocation.
	\item (C1') and (C2') for penalties conditions. 
\end{itemize}

To solve the problem in (\ref{eq:mindelta}), problem (\ref{eq:limy}) should be solved and thus constraints (C1') and (C2') should be respected. Further, constraints (C1)-(C4) should be respected to obtain an optimal solution to (\ref{eq:mindelta}). It is known that problem (\ref{eq:limy}) can be solved in an optimal way but tuning a parameter denoted as $V$ \cite{neely2012stability}. Thus, after verifying the constraints C1-C4, we can solve (\ref{eq:mindelta}) optimally as well. Our detailed solution is described in the sequel.

\section{Proposed Solution}

\subsection{Solution Overview}

Before going deep into the details of the solution, we provide the basic idea of the proposed resource provisioning approach. At the beginning of each resource provisioning cycle, we verify the constraints (C1)-(C4) related to the association, maximum resource capacities, type and requirement of each task. Then, we provide the resource provisioning through solving the optimization problem in (\ref{eq:limy}) by leveraging the Lyapunov optimization whenever a resource provisioning is needed under the constraints of (C1') and (C2'). Note here that we focus on the tasks that are to be processed at the edge of the network. At the EN side, the ENS checks the requirements for each task in terms of type and required amounts of resources. More details about our proposed approach are provided in the remainder of this section.

\subsection{Proposed Solution}

In order to solve problem in (\ref{eq:limy}), we consider $y_0^{min}$ as the minimum value of the penalty; in our case is the resource consumed over time slot $t$, and we put:

\begin{equation}
	y_k(t) = p_k(t)-p_k^{avg}
\end{equation}

$p_k(t)$ is the resource loss in the ED $k$ in the EN $i$ at the time slot $t$, and $p^{avg}_{k}$ is the average resource loss, and the constraint (C1') in (\ref{eq:limy}) holds if:

\begin{equation}
	\lim\sup\limits_{t\to\infty}\bar{y}_k(t)\leq p_k^{avg}
\end{equation}

Therefore, considering the Eq. (\ref{eq:Z}), we can rewrite $Z_k(t+1) = \max \left[Z_k(t) + y_k(t), 0\right]$ and we have $Z_k(\tau + 1) \geq Z_k(\tau) + y_k(\tau)$ for $\tau \in \{0, 1, ..., t-1\}$, due to requests arrival process which is i.i.d. Thus we can write \cite{neely2012stability}:

\begin{equation}
	\label{eq:Z-Z0}
	Z_k(t)-Z_k(0) \geq \mathlarger{\sum_{\tau=0}^{t-1}}y_k(\tau)
\end{equation}

we divide by $t$ in (\ref{eq:Z-Z0}), and we obtain:

\begin{equation}
	\frac{Z_k(t)}{t}-\frac{Z_k(0)}{t} \geq \frac{1}{t}\mathlarger{\sum_{\tau=0}^{t-1}}y_k(\tau)
\end{equation}

Clearly, $\frac{Z_k(t)}{t} \to 0$ when $Z_k(t)$ is stable, thus, $Q_i(t)$ is stable and (C2') is met.

Assuming that the queues are empty initially, we define $\theta(t) = [Qi(t), Z_k(t)]$ as the combination of the queue vectors, the Lyapunov function is expressed as follows:

\begin{equation}
	L(\theta(t)) = \frac{1}{2}\left[\mathlarger{\sum_{i\in\mathcal{E}}}Q_i(t)^2 + \mathlarger{\sum_{k\in\mathbb{D}_i}}Z_k(t)^2\right]
\end{equation}

The drift plus penalty process requires the minimization of $\mathbb{E}[\Delta(t)+V.y_k(t)~|~\mathcal{H}(t)]$, where $\Delta(t) = L(Q(t+1))-L(Q(t))$ and $V$ is a performance tradeoff parameter.\\
Let us assume that the functions of (\ref{eq:genfunc}) satisfy the following conditions for all values of $\alpha_i(t)$ and $\beta_i(t)$ :
\begin{equation}
	\begin{cases}
		a_i(t) \geq 0 \\
		b_i(t) \geq 0 \\
		y_0^{min}(t) \leq y_0(t) \leq y_0^{max}
	\end{cases}
\end{equation}
Where $y_0^{min}$ and $y_0^{max}$ are the maximum and minimum values of $y_0(t)$. Let $D \geq 0$ be a constant that for every resource allocation scheme $\alpha_i(t)$ based on values of $\beta_i(t)$ we have \cite{neely2012stability}:
\begin{equation}
	\begin{cases}
		\mathbb{E}\left[\hat{a_i}(\alpha_i(t), \beta_i(t))^4\right] \leq D \\
		\mathbb{E}\left[\hat{b_i}(\alpha_i(t), \beta_i(t))^4\right] \leq D \\
		\mathbb{E}\left[\hat{y_k}(\alpha_i(t), \beta_i(t))^4\right] \leq D
	\end{cases}
\end{equation}
Where $\mathbb{E}[.]$ represents the expectations taken, considering $\beta_i(t)$ and the decisions $\alpha_i(t)$.
The drift-plus-penalty expression for a finite constant $B > 0$, satisfies the following \cite{neely2012stability}:

\begin{equation}
	\label{eq:inneq}
	\begin{split}
		\mathbb{E}\left[\Delta(t)+V.y_0(t)|\mathcal{H}(t)\right] \leq \\
		B + V.\mathbb{E}\left[y_0(t)|\mathcal{H}(t)\right]+\\
		\mathlarger{\sum_{i\in\mathcal{E}}}\left(Q_i(t)\right)\mathbb{E}\left[a_i(t)-b_i(t)|\mathcal{H}(t)\right]+\\
		\mathlarger{\sum_{k\in\mathcal{D}_i}}\left(Z_k(t)\right)\mathbb{E}\left[y_k(t)|\mathcal{H}(t)\right]
	\end{split}
\end{equation}

The right-hand-side of Eq. (\ref{eq:inneq}) is minimized when choosing a minimal value of $\alpha_i(t)$ corresponding to the states of the $Q_i(t)$ and $Z_k(t)$ and the state of resource at EN $i$, $\beta_i(t)$:

\begin{equation}
	\label{eq:v}
	V.y_0(t) + \mathlarger{\sum_{i\in\mathcal{E}}}Q_i(t)\left[a_i(t) - b_i(t)\right] + \mathlarger{\sum_{k\in\mathcal{D}_i}} \left[Z_k(t)y_k(t)\right] \leq C
\end{equation}

$Q_i(t)$ and $Z_k(t)$ are updated according to (\ref{eq:Q}) and (\ref{eq:Z}), with the existence of a given constant $C \geq 0$, optimal value of $\alpha_i(t)$ is chosen when $C$ is small as possible, and:

\begin{equation}
	\begin{split}
		\mathrm{V}.y_0(t) + \mathlarger{\sum_{i\in\mathcal{E}}}Q_i(t)\left[a_i(t) - b_i(t)\right] + \mathlarger{\sum_{k\in\mathcal{D}_i}}\left[Z_k(t)y_k(t)\right] \\ \leq C + \\\inf_{\alpha_i(t)}[V.y_0(t) +  \mathlarger{\sum_{i\in\mathcal{E}}}Q_i(t)\left[a_i(t) - b_i(t)\right] + \\ \mathlarger{\sum_{k\in\mathcal{D}_i}}\left[Z_k(t)y_k(t)\right]]
	\end{split}
\end{equation}
In case $C = 0$, the exactly minimum value of $\alpha_i(t)$ is reached, thus the optimal resource allocation scheme is obtained.

The resource representation allows characterizing the exact capabilities of the EDs precisely and uniformly. The resource representation allows the ENS to get information about the EDs resource status, which correspond to the values of $\beta_i(t)$ in our proposed approach in the previous section. Each ED is aware of its purpose (the types of operations that could be handled), its capacities and available resources. For this purpose, each ED exposes its available resource to the ENS through the MEC API standard of ETSI \cite{ETSIGRMEC027,GSMEC010,GSMEC011,GSMEC021,ETSI5G}.

\begin{figure}[!t]
	\centering
	\includegraphics[width=.7\linewidth]{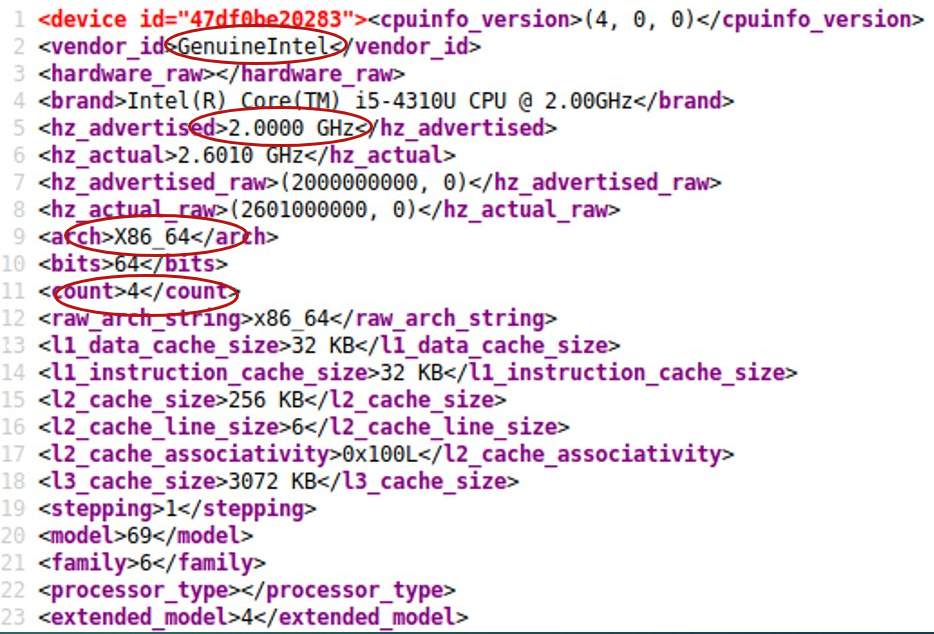}
	\caption{Example of output for an ED exposing information about its CPU}
	\label{fig:exmpl}
\end{figure}

The resources are exposed in XML format to the supervising entity. In order to get information about the physical resource, each ED uses low-level operations and commands including \textit{CPUID} to get the CPU properties such as the architecture, number of cores and the frequency. The ED, depending on its OS, uses commands such as \textit{cpuinfo} and \textit{meminfo} for Linux based machines to obtain the current state of the CPU usage and memory respectively, \textit{wmic} for Windows-based EDs. We proceeded with the same to cover the majority of the used OS in the EDs. Collecting all these commands within the same program that, depending on the ED used OS, collects the information about the EDs and their available resources. Figure \ref{fig:exmpl} shows an example of an ED’s response when the supervisor requested information about the CPU resource. The output is an XML response with all the different information about the CPU such as the family, the frequency, number of cores and the architecture. Also, XML can be parsed by almost all EDs, which makes the information about the resource easier and more understandable to the ENS.

Algorithm \ref{algo:resalloc} is performed in order to define the optimal allocation scheme for determining the adequate EDs that will participate in processing the incoming request. The time complexity of the proposed Lyapunov-based optimization algorithm is around $O(N^2)$ \cite{neely2012stability}, which ensures an low complexity of our proposed algorithm in Algorithm 1, since it is the only heavy task to perform by the ENS.

The frequency of checking ED resources (through resource representation model) depends essentially on the frequency of reallocating the resources. More precisely, the resource's status is checked every time the resource allocation is needed instead of being checked at each given time slot as in \cite{8962353, liu2019dynamic, 8683499, guo2019mobile, 8815852, liao2019learning, zhou2019computation, gu2019task}. We could solve this problem either statically or dynamically. In the static scheme, the problem is solved by fixing a time interval on which we perform the resource's checking, which is not optimal since the resource reallocation interval on a given EN is longer under a low load and shorter under a high load. We can solve the problem of reallocation resources dynamically by monitoring the load on devices \cite{lin2011threshold}. Below, we consider that the moment of checking resource's status is the moment of reallocating resources. In fact, reallocating virtual resources for edge computing applications costs additional computational resource. Thus, overly frequent reallocation of resources might decrease the efficiency of the EN resources usage. Inspired from the the work in \cite{lin2011threshold}, we define the workload of an ED for a given resource $\sim$ (where $\sim~\in \{\mathcal{P}, \mathcal{S}, \mathcal{N}, \mathcal{M}\}$) as follows:
\begin{equation}
	\label{eq:load}
	L_{(i,j,k)}^{r_{(i,j)}^\sim}=r_{(i,j)}^\sim-\sum_{k=1}^{|C_{(i,j)}|}\Xi_{(i,j)}^{\sim,k}
\end{equation}

\begin{algorithm}[!t]
	\caption{Resource allocation based on resource representation and Lyapunov optimization (LRR)}
	\label{algo:resalloc}
	\begin{algorithmic}[1]
		\State \textbf{Input:} incoming requests $\varrho_{(i,j)}$
		\State \textbf{Init:} place $\varrho_{(i,j)}$ at the queue
		\State get $D_{u_k}$ and $s_{(i,j)}$
		\If {\textit{free}($r_{(i,j)}^\sim$)}
		\State Allocate resource at ED $d_{(i,j)}$
		\State Instantiate container $c^{k}_{(i,j)}$ for the requested service with respect to constraint in equation (2)
		\State Destroy the $c^{k}_{(i,j)}$ after task completion
		\Else
		\State Find $\alpha_i$ from the problem in (\ref{eq:limy}) determine the candidate ED with respect to $T_{out}$
		\State Allocate the adequate amount of resource on each selected ED from (\ref{eq:limy})
		\State Distribute the subtasks to the ED
		\State Create containers on each ED
		\State Destroy containers after all subtasks are finished
		\EndIf
		\State Wait for the next request
	\end{algorithmic}
\end{algorithm}

If we assume that the maximum workload of an ED is given by $L^{max}_{(i,j,k)}$, we define the normal workload $L^{norm}_{(i,j,k)}$ as follow:

\begin{equation}
	L_{(i,j,k)}^{norm}=L_{(i,j,k)}^{max}\times rate_{norm} \text{ with } rate_{norm}\in[0,1]
\end{equation}
Finally, we can define $L_{(i,j,k)}^{th}$ as the threshold that triggers the resource reallocation process as :
\begin{equation}
	L_{(i,j,k)}^{th}=rate_{th}\times(1-rate_{norm})\times K\times L_{(i,j,k)}^{max}
\end{equation}
Where $K$ represents the number of container on execution on the ED $j$. A resource reallocation takes place when the value of $L^{th}_{(i,j,k)}$ changes.

\begin{algorithm}[!b]
	\caption{Resource Reallocation Calculation}
	\label{algo:frequ}
	\begin{algorithmic}[1]
		
		\State Use (29) and (30) to calculate $L^{norm}_{(i,j,k)}$ , $L^{th}_{(i,j,k)}$ and $L^{max}_{(i,j,k)}$
		\For{\textbf{each} $d(i,j)$ in $D_i$}
		\If {$L^{curr}_{(i,j,k)} > L^{norm}_{(i,j,k)} + L^{th}_{(i,j,k)}$}
		\State send resource information request to $d(i,j)$
		\EndIf
		\EndFor
	\end{algorithmic}
\end{algorithm}

The time complexity of algorithm (\ref{algo:frequ}) depends on the number of ED in the EN. Since the ENS has the information about each ED at the beginning of the network operations, information about the loads are available, since the ENS is aware of the tasks' sizes and the capabilities of each ED. Moreover, the number of the resource allocation scheme changing in some cases remains unchanged due to the availability of resource on a given ED, which ensures the optimality of the resources used to compute the optimal allocation scheme $\alpha_i(t)$ \cite{lin2011threshold}. In other words, the frequency of performing resource allocation is lower when compared to other schemes that perform the same operation not only at each request arrival but at each time slot \cite{8962353, liu2019dynamic, 8683499, guo2019mobile, 8815852, liao2019learning, zhou2019computation, gu2019task}.

\section{Simulation Results}

\begin{figure*}[!t]
	\centering
	\includegraphics[width=.8\linewidth]{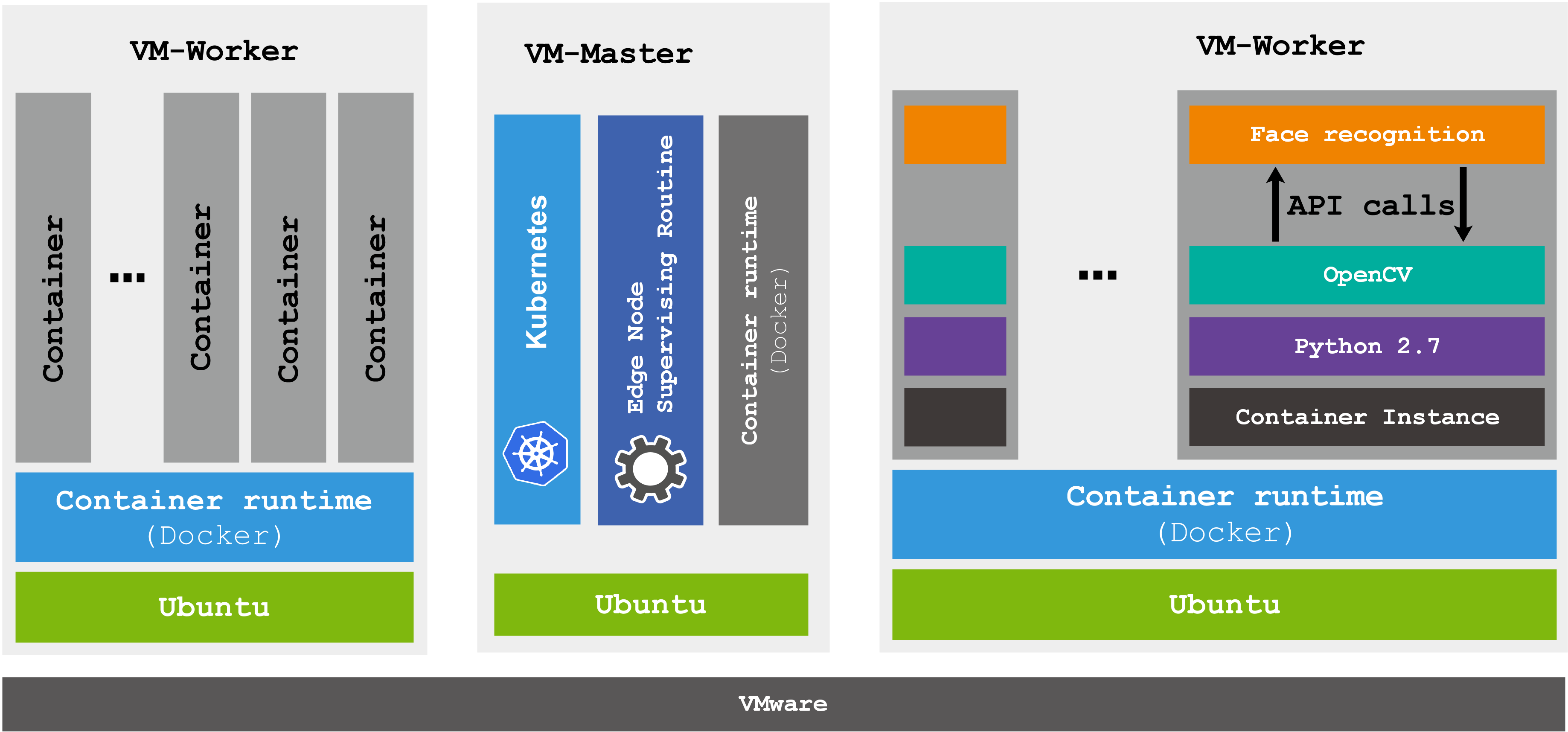}
	\caption{Testbed components; The VM-Workers (EDs) on left and right and the VM-Master (ENS) on the center.}
	\label{fig:simulation-scheme}
\end{figure*}

\begin{figure*}[!t]
	\centering
	\includegraphics[width=\linewidth]{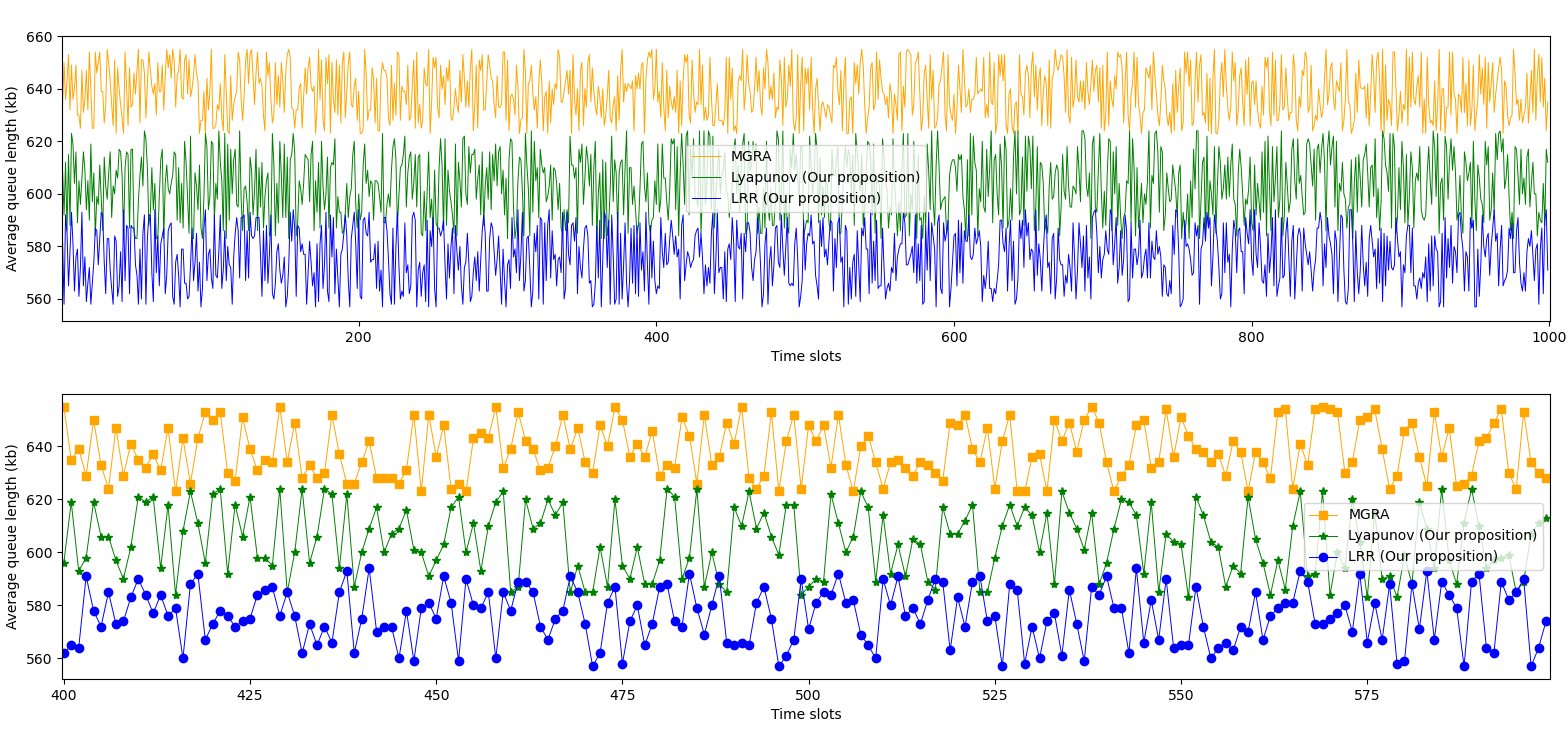}
	\caption{Average queues sizes over time}
	\label{fig:fig4}
\end{figure*}

We evaluated our proposed resource provisioning approach through extensive simulations. The experimental testbed was inspired from the work of \cite{rimal2018experimental}. For simplicity,  we used the same OS (Linux) and we deployed Docker on these VMs, then we used Kubernetes as container management engine. We then configured the VMs to act as a swarm from the point of view of Docker. We deployed our resource representation component over these VMs, and we deployed  our resource allocation algorithm on the master device (the ENS). Moreover, we tested the resource representation component on different versions of OSs and different hardware architectures (x86, x64 and ARM). However, the testbed with heterogeneous configuration is yet to be investigated in a future work. The experimental simulations consist of an EN with 3 EDs. As illustrated of Fig. \ref{fig:simulation-scheme}, with the ENS (master) is at the middle having the role of supervising the EN, receiving the request and executing the resource provisioning schemes. The central ED in the Fig. \ref{fig:simulation-scheme} is considered to host the master of the EN, thus, the ENS. The other EDs are the workers. The ENS is responsible for checking the EDs resource information, performing the resource provisioning (computing the optimal scheme and values of $\alpha_i(t)$) and forwarding the requests to the adequate nodes based on the $\alpha_i(t)$ values). We tested the proposed approach by implementing a face recognition based on OpenCV framework \cite{kaehler2016learning}. We simulated the proposed approach using a setup that is closely equal to the experimental one, having a power of processing of $2.5Ghz$, a memory variation between $\{2-4\}$ Gb and a storage between $\{20-30\}$ Gb. We also considered the $V$ parameter ranging in $0-100$. To put the testbed to work, we considered the following scenario, in which we used a face recognition application inspired from \cite{muslim2017face}. The user's device (smartphone for instance) sends the images to the EN for recognition. The face recognition classifier is already trained in the EN. After receiving the image, the EN solves the optimization problem and select the EDs in which the face detection should take place. The EDs have different capabilities and configurations as described in the table follows:

\begin{table}[h]
	\centering
	\caption{Experimental Setup}
	\begin{tabular}{|l|l|l|l|l|}
		\hline
		\multicolumn{2}{|l|}{Nodes/Resource capabilities} & Memory & Processing & Storage \\
				\hline
		\multirow{3}{*}{EN}            & ED1            & 2Gb    & 2 Cores    & 20Gb    \\
		& ED2            & 2 Gb   & 4 Cores    & 30Gb    \\
		& ED3            & 4Gb    & 2 Cores    & 20Gb   \\
				\hline
	\end{tabular}
\end{table}

\begin{figure}[!t]
	\centering
	\includegraphics[width=.8\linewidth]{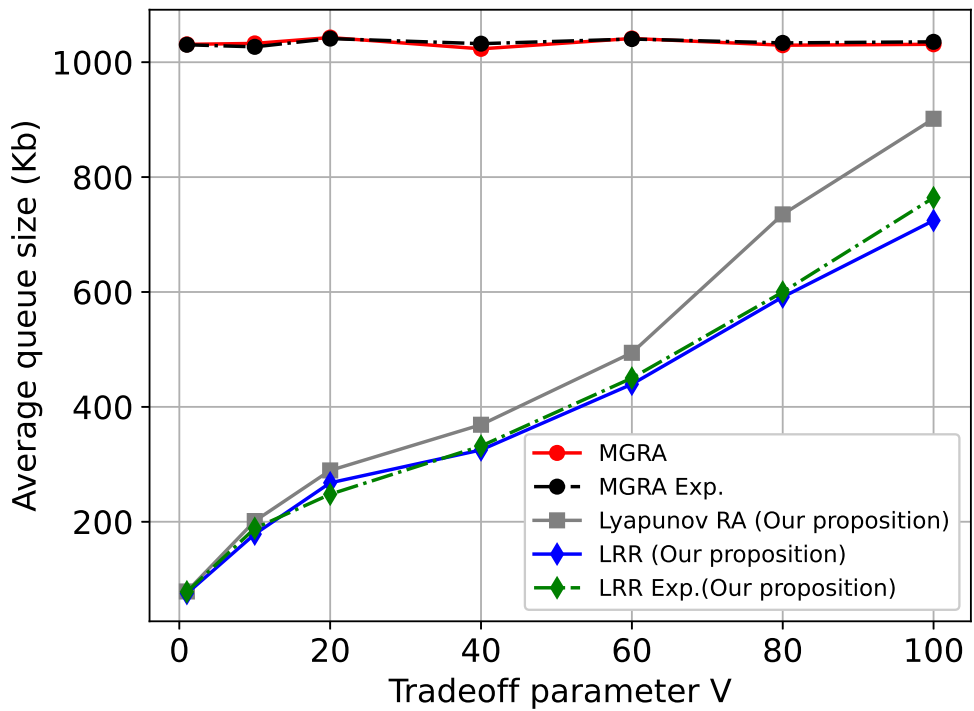}
	\caption{Average queue size evolution in function of the parameter V}
	\label{fig:fig3}
\end{figure}

We compare our approach to the MGRA benchmark approach \cite{guo2019mobile} both theoretically and in simulations. We also compare our approach when using both resource representation and Lyapunov optimization (LRR in Fig. 4-11) with using the proposed Lyapunov framework alone. Fig. \ref{fig:fig3} illustrates the average queue size evolution in time. The queue size is minimized in LRR compared to MGRA. The figures also show that using resource representation in our approach has a positive effect in minimizing the queue size compared to using our proposed Lyapunov framework alone. In addition, the mechanism we used is lightweight and could be executed within a tiny shred of resources.

In order to test the impact of the parameter $V$ of Eq. (\ref{eq:v}) on the overall system performance, we evaluated the average queue size under the different values of $V$ to find the optimal configuration of the performance-delay tradeoff. As a reminder the $V$ parameter represents the tradeoff between the performance and the allowable latency.

\begin{figure}[!t]
	\centering
	\includegraphics[width=.8\linewidth]{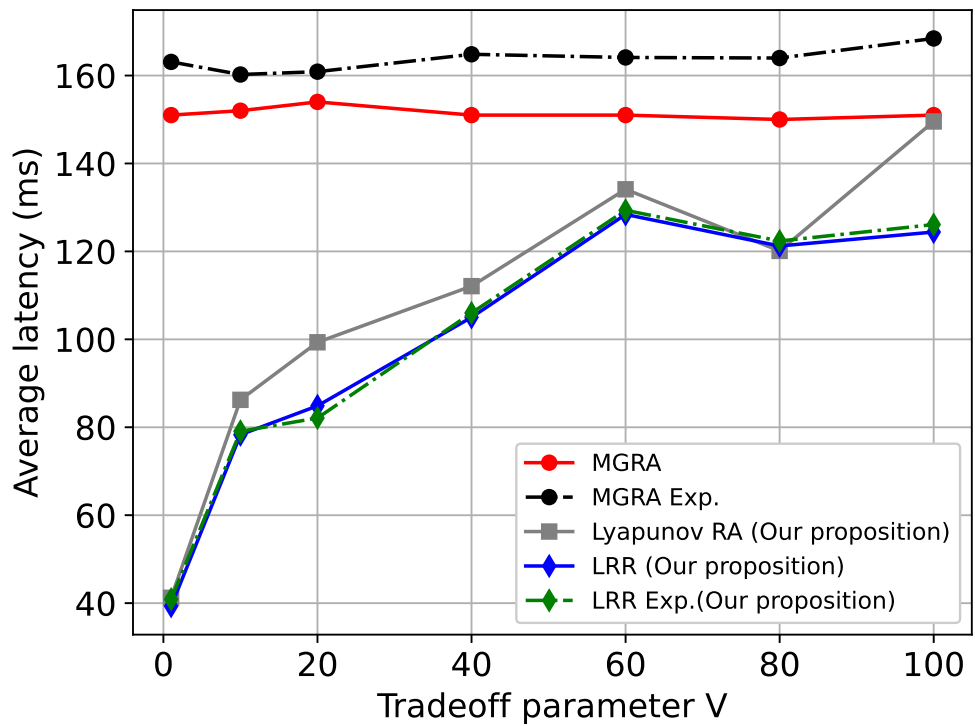}
	\caption{Average latency evolution in function of the parameter V}
		\label{fig:fig2}
\end{figure}

We illustrate the evolution of the queue under the variation of the $V$ parameter. Fig. \ref{fig:fig3} which shows that a lower value of $V$ means that the user is interested in being served within a low delay. In the MGRA scheme, there is no consideration to such parameter, which causes the constant behavior of the average queue size. Fig. \ref{fig:fig3} shows that the queue size of is less congested when using our Lyapunov framework and even less congested when using LRR. Even for high values of $V$ , the queue is far from reaching the queue size of the benchmark MGRA scheme.

\begin{figure}[!t]
	\centering
	\includegraphics[width=.8\linewidth]{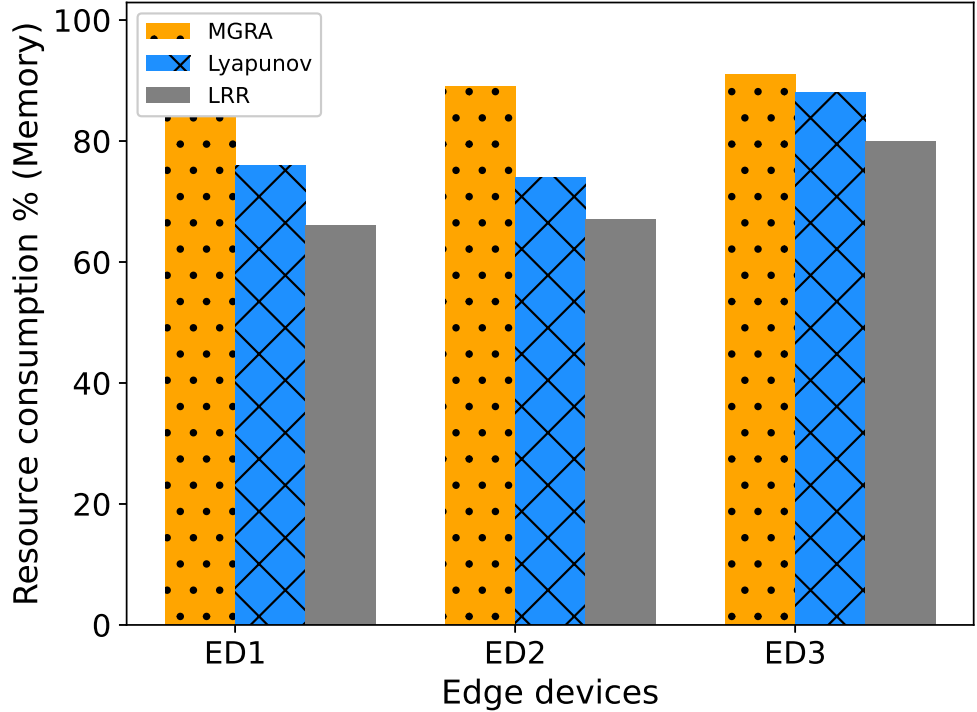}
	\caption{The average network interface utilization on each ED of the simulation setup}
	\label{fig:net}
\end{figure}
We evaluated the impact of the parameter $V$ on the average latency and we conclude that a higher value of $V$ implies a higher latency, which is also very clear from the figure \ref{fig:fig4}. Our proposed approach shows an improvement in terms of latency. Fig. \ref{fig:fig1} illustrates the average latency evolution with the queue size under the different schemes of resource provisioning. The average latency represents the overall delay spent from the moment of requesting a task processing, to selecting the adequate ED to process the task and then receiving the results. Our proposed LRR approach outperforms the MGRA approach.

Our approach achieves the lowest latency with the evolution of the queue size, and with further examination, Fig. \ref{fig:fig1} shows that LRR improves the latency up to 40\%. This can be explained by the fact that the ENS is all the time aware of the available resources at the ED and their capabilities. Therefore, based on the resource information, the ENS allocates the resources on the adequate EDs, which distribute the tasks in an adequate way that guarantees a lower latency for the users and resource consumption at the ED following the Lyapunov optimization framework.

\begin{figure}[!b]
	\centering
	\includegraphics[width=.8\linewidth]{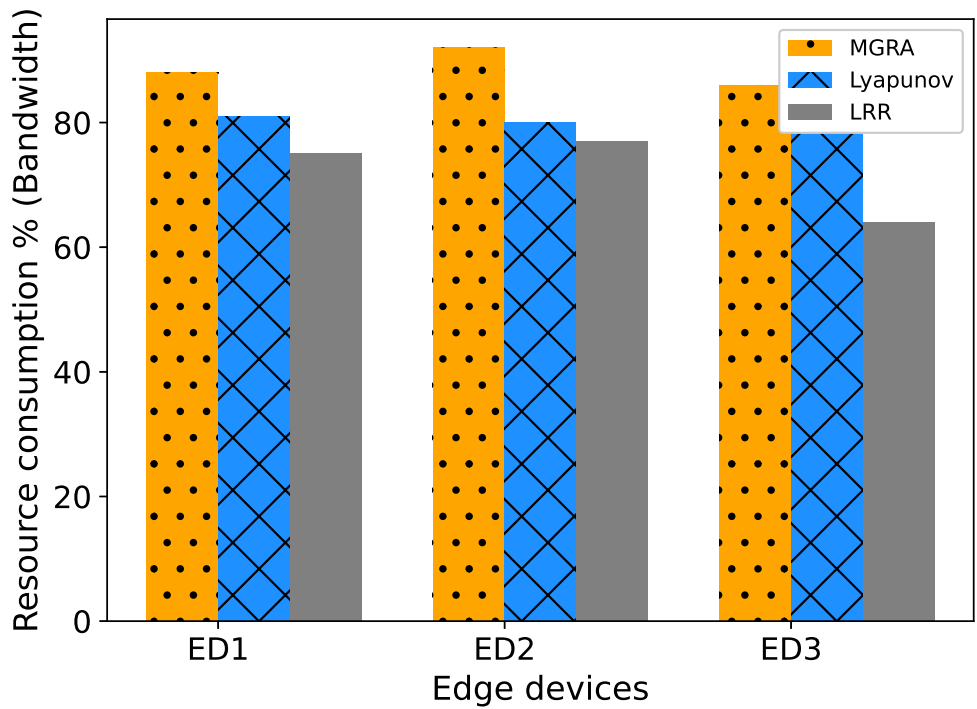}
	\caption{The average storage utilization of each ED from the setup}
	\label{fig:memo}
\end{figure}

\begin{figure}[!b]
	\centering
	\includegraphics[width=.8\linewidth]{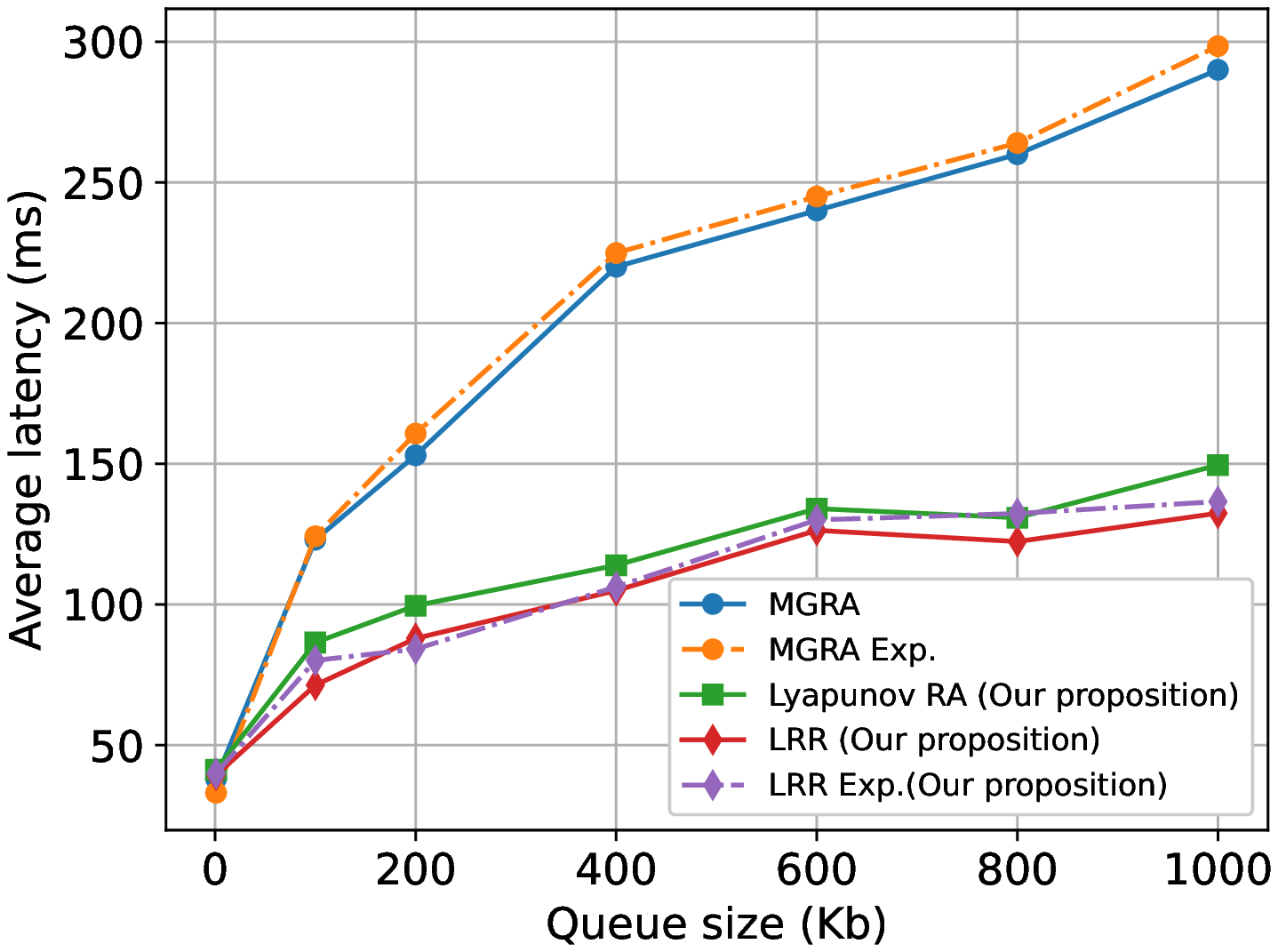}
	\caption{Latency evolution over the queue size}
	\label{fig:fig1}
\end{figure}

\begin{figure}[!t]
	\centering
	\includegraphics[width=.8\linewidth]{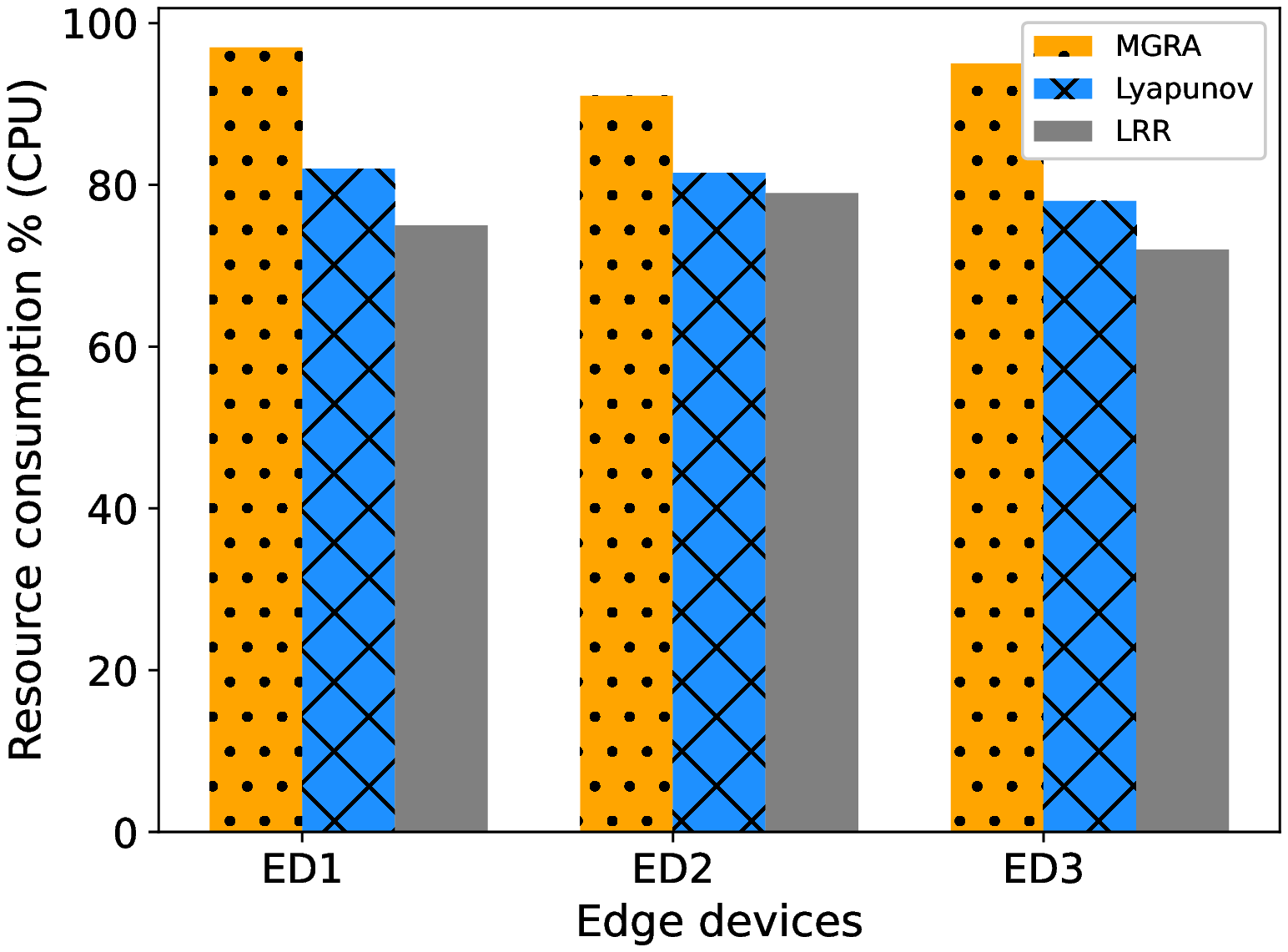}
	\caption{Average CPU utilization at the EDs.}
	\label{fig:cpu}
\end{figure}

\begin{figure}[!t]
	\centering
	\includegraphics[width=.8\linewidth]{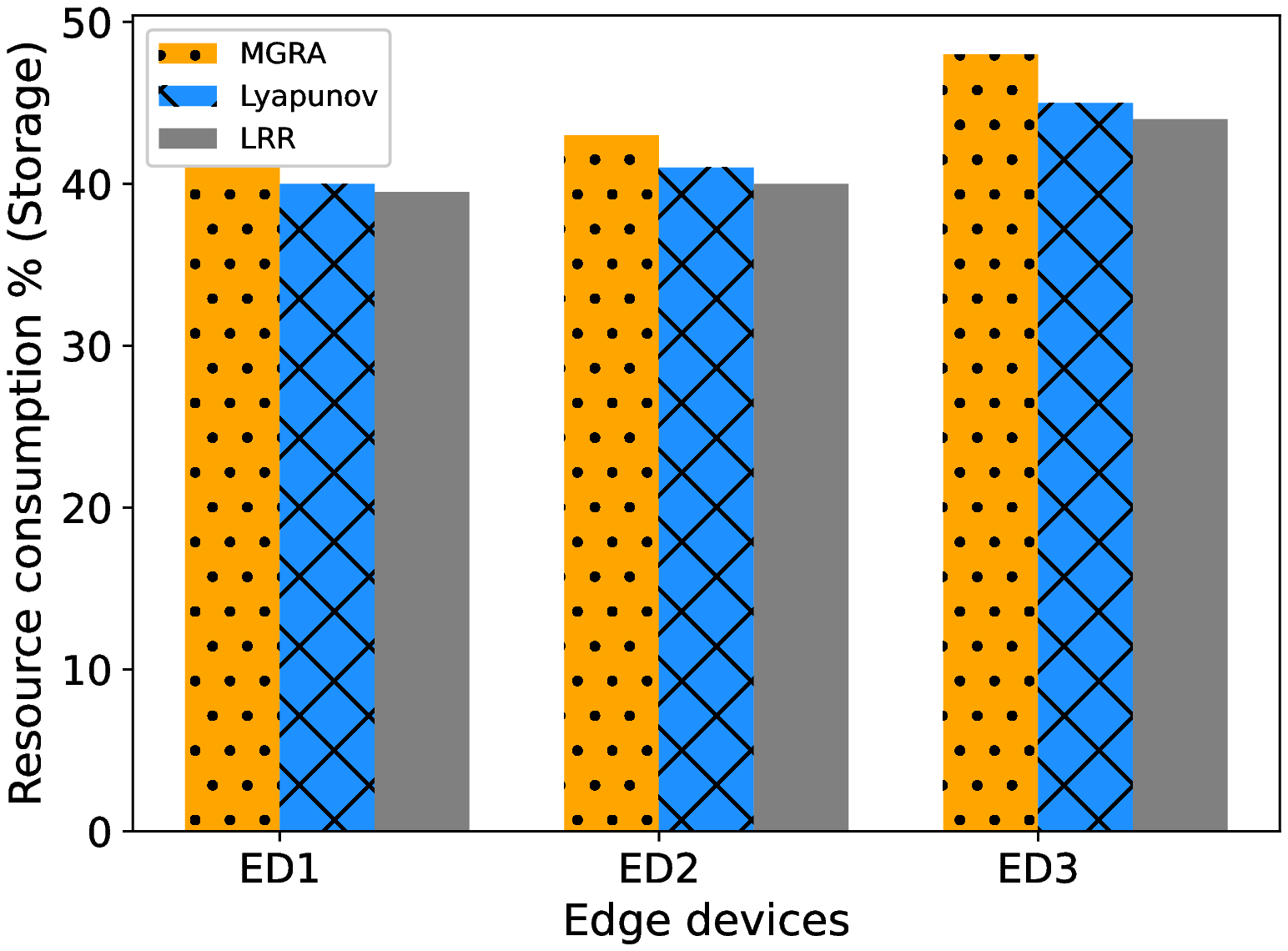}
	\caption{The average storage utilization of each ED from the setup}
	\label{fig:strg}
\end{figure}

The results for our study of the testbed ED resources behavior under the different resource provisioning schemes are illustrated in Fig. (\ref{fig:fig4}-\ref{fig:strg}). Our proposed LRR scheme shows a significant improvement in terms of CPU consumption as shown in Fig. \ref{fig:cpu}. This enhancement is due to the ENS awareness of the resource status of each ED. When adopting the benchmark MGRA approach, a portion of the resources from the devices are used to compute the optimal matching between the users and the EN which takes almost $O(N^2)$ compared to Lyapunov which takes only $O(\frac{1}{N2})$. In some cases, the MGRA benchmark approach which is based on a matching game, takes long delay to reach a stable distribution, which may cause higher consumption of resources. In LRR, the resource representation guarantees the information about the ENs resources. Using Lyapunov optimization for addressing the queue’s congestion offers better resources utilization on one hand. On the other hand, it is mathematically proved that an optimal resource allocation scheme is always guaranteed within a reasonable delay. Fig. \ref{fig:cpu} shows that LRR guarantees a lower resource consumption in terms of CPU compared to the benchmark approach; specifically approximately a 11\% lower consumption on ED1 and ED2, and approximately a 18\% lower on ED3.

Fig. \ref{fig:strg} illustrates the storage consumption under the MGRA, and our approach using Lyapunov with and without considering the resource representation. Fig. \ref{fig:strg} shows that the different schemes have almost the same performance for ED1 and ED2. However, in ED3, LRR significantly improves the performance of storage use by almost 25\% compared to MGRA, due to the fact that in our proposed scheme we used containers as a virtualization technology instead of VMs used in the MGRA approach.

The network performance is illustrated in Fig. \ref{fig:memo}, in which the Lyapunov-based approach shows an enhancement of the bandwidth consumption as it was already been discussed in previous papers such as the works in \cite{8962353,liu2019dynamic}. In addition, the resource representation gives better results, 15\% lower than MGRA and 6\% lower than the Lyapunov-only based approach.

Fig. \ref{fig:net} illustrates the memory consumption on the different EDs of our testbed. The results show that LRR uses 14\% less resources compared to the benchmark scheme. This result could be explained by the fact that in a matching game approach, EDs are not aware of what the other devices have in terms of capabilities. Also, the matching game is distributed and the discovery phase could take much more time. Our LRR approach which makes the information about available resources accessible to the ENS while using a Lyapunov approach makes the resource allocation more efficient.

\begin{figure}[!t]
	\centering
	\includegraphics[width=.8\linewidth]{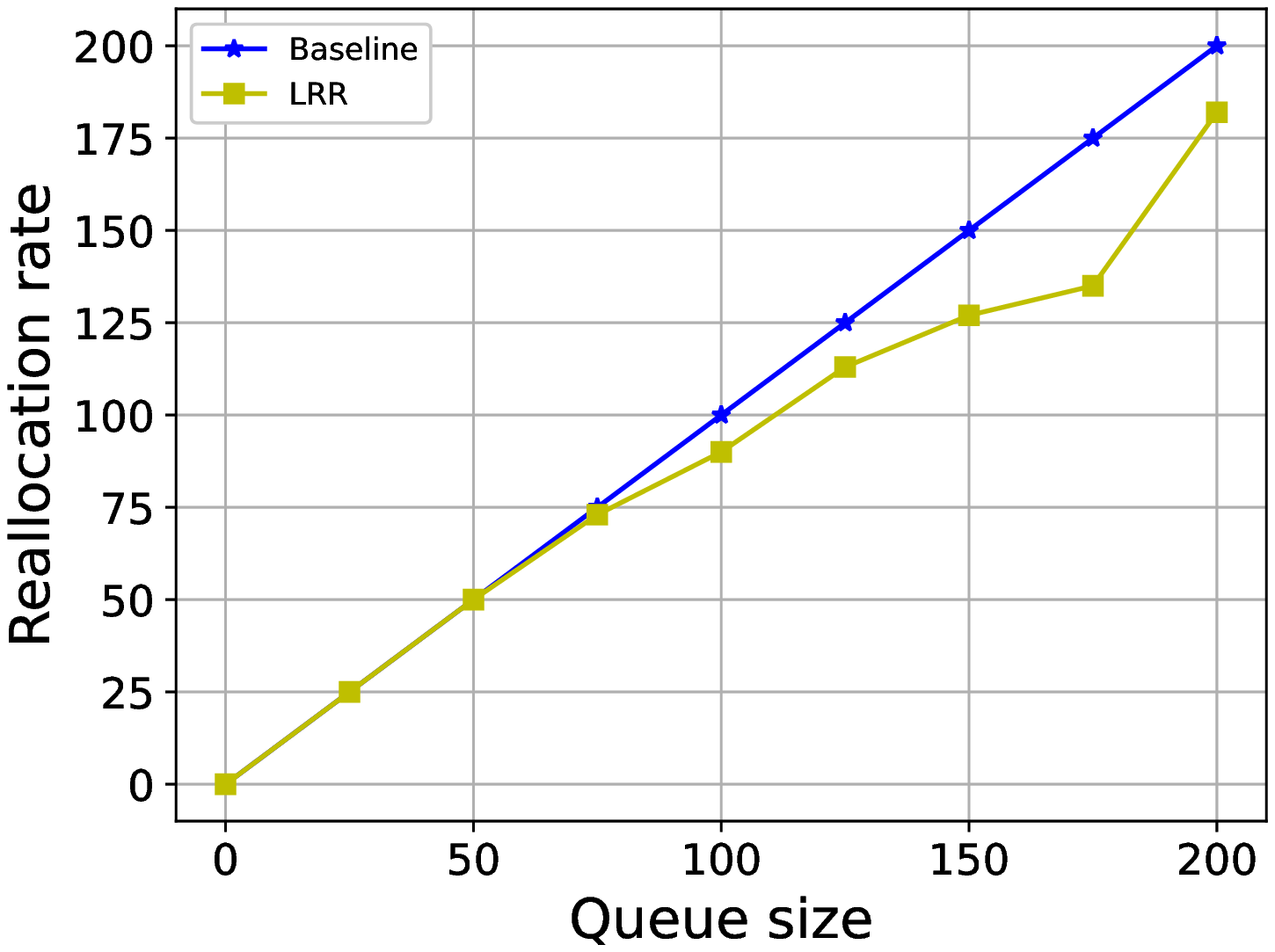}
	\caption{The reallocation frequency in function of the queue size}
	\label{fig:frequ}
\end{figure}

Fig. \ref{fig:frequ} illustrates the evolution of the reallocation frequency in LRR compared to a baseline scheme. The baseline scheme referred to in Fig. \ref{fig:frequ}, is used in several previous works \cite{8962353, liu2019dynamic, 8683499, guo2019mobile, 8815852, liao2019learning, zhou2019computation, gu2019task, lin2011threshold} and consists in allocating resources at each time slot where a request is received. However, adopting such policy may make the EN reallocate resources continuously at each node to satisfy the requests, which may have a direct impact on the latency. In LRR, the ENs have sufficient information about each ED's capabilities and up-to-date resource states, making it easier to allocate resources on nodes with available resource without the need of recomputing the allocation scheme.

\section{Related Works}

Several previous works investigated the resource allocation and management from the edge computing perspective. Most of these works, such as in \cite{8962353, liu2019dynamic, 8683499, guo2019mobile, 8815852, liao2019learning, zhou2019computation, gu2019task} investigate the resource allocation problem for only one or two types of resources, and very few investigated the problem of resource allocation with more than two types of resources \cite{tocze2018taxonomy}; mainly the computational and networking resources. In this paper, we propose a resource allocation scheme for four kind of resources including processing, storage, memory and networking through resource representation. Resource representation allows the ENS to acquire information about all the available resources at each device in the EN. Some of the previous works did address the resource allocation problem using the Lyapunov optimization framework for minimizing the queuing time \cite{8962353, liu2019dynamic, 8683499, liao2019learning, liu2017fronthaul}. However, these works focus on few types of resource and the allocation cycle takes place at each time slot, which may lead in some cases to compute the optimal allocation scheme even when the current scheme can deal with incoming requests. Therefore, in our proposed approach the resource allocation scheme is only performed when the incoming request  requires a re-computation of the resource allocation scheme. Other studies such as in \cite{guo2019mobile, 8815852, liao2019learning, zhou2019computation, gu2019task} used theoretical game-based approaches such as matching games and coalitional games.

In some papers, authors propose a combination of different techniques such as in \cite{8962353, liu2019dynamic, 8683499}, combining Lyapunov optimization and matching game frameworks or combining multiple optimization framework such as in \cite{liao2019learning}. In \cite{8962353} the authors proposed a mixed-integer nonlinear programming problem to optimize the task offloading, computation scheduling and the radio resources. The problem was relaxed to subproblems by leveraging the Lyapunov optimization framework, then proposing a convex decomposition approach and matching game to solve the subproblems. In their proposed algorithm for the optimal task offloading, decision and resource allocation is taken at each time slot, which may lead in some scenarios to changing the overall decision to offload a small portion of tasks. In the study in \cite{liu2019dynamic} the authors proposed a user-server association scheme that takes the channel quality in account in addition to the computational capabilities and workloads of the servers. The authors also used a Lyapunov optimization and a matching game to propose a dynamic task offloading and resource allocation policy. The authors of \cite{8683499} proposed a resource allocation scheme for computational and communication resources jointly with users' access point association. The proposed solution in \cite{8683499} is based also on a Lyapunov framework to stabilize the queue and a matching game theory to associate the users to the access points. From the game theoretical perspective, the resource allocation scheme proposed in \cite{guo2019mobile} is based on a joint Stackelberg game to efficiently allocate computing resources to devices, and a one-to-many matching game to match the users to access points. The approach is similar to the one in \cite{8815852}, where instead of using different game theory frameworks, the authors proposed two-tiers matching game to optimize computing resource allocation and pricing, under limited computing and communication resources constraints. The first tier aims at associating users and small base stations with the goal of maximizing the social welfare. The second tier aims at achieving the collaboration between the different small base stations in order to ensure an efficient computing resources consumption. In some papers such as in \cite{liao2019learning}, the authors investigated the optimization of the network resources, more specifically the channel selection which is critical to ensure a reliable task allocation. The authors proposed the combination of three optimization tools to optimize the long-term throughput under energy cost and service reliability constraints. The authors also proposed a learning-based channel selection with service reliability, energy, backlog, and conflict awareness through with a Lyapunov optimization framework to optimize the strategy to allocate the channel, and a matching game approach for the channel selection. However, in these works, the aspect of resource is used in an abstract manner, and the proposed approaches does not put assumptions about either the heterogeneity of the equipment used, neither the interaction between the equipment.

\section{Conclusion}
In this paper, we investigated resource provisioning at the edge of the network under latency and resource consumption constraints. In order to reduce the latency, we studied the experienced delays at the different levels of the considered architecture. Precisely, we studied the queue dynamic at the EN by leveraging a Lyapunov optimization framework. In order to perform efficient resource provisioning, we proposed a resource representation for EDs. The resource representation allows the exposition of EDs resource information (processing, storage, memory and networking) at any time through the ETSI standard for edge computing applications. The ENS uses the gathered information on available resources to define the optimal resource provisioning scheme based on drift-plus penalty of the Lyapunov optimization framework. We also studied the frequency of resource reallocation and we proposed an algorithm based on the workload at each ED of the EN to reduce the number of times at which the ENS performs a resource reallocation operation. Moreover, we performed extensive theoretical and experimental simulations to prove the effectiveness of our proposed approaches. The numerical results have shown that our proposed approach outperforms the benchmark approach in terms of latency which drops up to 25\% in some cases, and up to 40\% in terms of lower resource consumption.
\bibliographystyle{IEEEtran}
\bibliography{refdb}

\end{document}